\documentclass{lhep}       
\journal{LHEP}
\vol{xx}
\jyear{2018}
\pages{xxx} 
\received{xx January 2018}
\published{xx March 2018}

\def\be{\begin{equation}}
\def\ee{\end{equation}}
\def\bea{\begin{eqnarray}}
\def\eea{\end{eqnarray}}

\usepackage{amsmath}
\usepackage{xcolor}
\usepackage{amssymb}
\usepackage{hyperref}
\usepackage{slashed}
\usepackage{mathrsfs}
\usepackage{braket}
\usepackage{graphicx}

\begin{document}
\title{Neutrino oscillations induced by chiral torsion}
\author{Riya Barick, Indrajit Ghose, Amitabha Lahiri}
\address{Satyendra Nath Bose National Centre for Basic Sciences, JD Block, Sector - III, Salt Lake, Kolkata 700106, INDIA}

\begin{abstract}
Neutrino mixing is caused by the fact that neutrino flavors are not eigenstates of the free Hamiltonian. This causes oscillations among different neutrino flavors. When neutrinos pass through a medium, weak interactions produce different effective masses for neutrinos of different flavors, leading to a modification of the mixing parameters. In curved spacetime there is an additional contribution to neutrino Hamiltonian from a torsion-induced four-fermion interaction, which also causes  neutrino mixing while propagating through fermionic matter. We provide an outline of the calculation of this effect on neutrino oscillation. 

\vspace{0.3cm}
\textbf{Keywords} \hspace{0.5cm} Chiral Torsion, NSI, Neutrino Mixing.
\end{abstract}
\maketitle
\newpage
\section{Introduction}
\label{sec:intro}
The Standard Model says that neutrinos are massless and purely left-handed, described by two component Weyl spinors. 
On the other hand, the disappearance or appearance of flavors among neutrinos from various sources -- top of the atmosphere~\cite{Super-Kamiokande:1998kpq}, the Sun~\cite{SNO:2002tuh}, reactors~\cite{KamLAND:2002uet}, accelerators~\cite{T2K:2011ypd} and in many other experiments~\cite{K2K:2002icj,MINOS:2006foh} point to neutrino flavor mixing and oscillation, which is explained by the existence of neutrino masses. This clearly indicates the presence of physics beyond the Standard Model. There are many ways to extend the Standard Model to include massive neutrinos~\cite{Mohapatra:1998rq}. 
Different mechanisms will have different consequences for neutrino oscillations~\cite{Masud:2021ves,Chatterjee:2015gta}. 

One kind of BSM physics that is not usually considered is the interaction of fermions with spacetime geometry. Fermions in a curved spacetime generate a spacetime torsion, which is non-dynamical and can thus be integrated out, leaving a quartic interaction term~\cite{Chakrabarty:2019cau, Lahiri:2020mst}.  Although gravity is a very weak force, the freedom to choose coupling constants means that the contribution of spacetime geometry on neutrino oscillations may not be negligible. In what follows, we present a brief outline of a calculation for three flavors of neutrinos~\footnote{Based on talk delivered by {Riya Barick} at the International Conference on Neutrinos and Dark Matter (NuDM-2022), Sharm El-Sheikh, Egypt, 25-28 September, 2022.}.

The layout of the paper is as follows. In Sec.~\ref{sec:spacetime} we provide a very brief review of how fermions behave in spacetime and then write how an effective quartic interaction arises. In Sec.~\ref{sec:2flavor} we calculate the modified mixing matrix for two neutrino flavors, and then we do the same calculations for three neutrino flavors in Sec.~ \ref{sec:3flavor}. 

\section{Fermions in spacetime}
\label{sec:spacetime}
Usually the geometry of spacetime is affected in presence of matter. The effect of matter on bosons can be neglected for small curvature, as in most of the physical phenomena. But the case is different for a fermionic field. When a fermion passes through a fermionic matter, a four-fermi interaction happens there, which couples with fermions with different coupling constants which are fixed by experiments. 

Torsion $ \Lambda_{\mu}{}^{ab} $, is added to the Levi-Civita connection $ \omega_{\mu}{}^{ab} $, in the 1st order formulation of Gravity, which uses tetrad fields, through the relation~\cite{Cartan1922, Poplawski:2009fb}
\begin{equation}\label{split}
A_{\mu}{}^{ab}=\omega_{\mu}{}^{ab} +\Lambda_{\mu}{}^{ab}  \,,
\end{equation}
where $ A_{\mu}{}^{ab} $ is called spin connection. If we assume that $ \Lambda $ couples chirally to fermions, the equation of motion  for $ \Lambda $ is
\begin{equation}\label{chiral.torsion}
	\Lambda_{\mu}{}^{ab} = \frac{\kappa}{4}\epsilon^{abcd}e_{c\mu} \sum\limits_i \left(-\lambda^i_{L}\bar{\psi}_{iL}\gamma_d \psi_{iL} + \lambda^i_{R}\bar{\psi}_{iR}\gamma_d \psi_{iR}\right)\,,
\end{equation}
which is clearly also its solution. Here  $\kappa = 8\pi G$\,, while $ e^a_{\mu} $ are the tetrad fields and $ e^\mu_a $ are their inverse fields, defined by $\eta_{ab}e^a_{\mu}e^b_{\nu} = g_{\mu\nu}\,.$ The tetrads can be combined into one 4$\times$ 4 matrix, has determinant equal to the square root of the metric determinant, $|e| = \sqrt{|g|}$\,. Then we can put $\Lambda$ back into the action to get an effective quartic interaction term~\cite{Chakrabarty:2019cau}
\begin{equation}\label{4fermi}
	-\frac{1}{2}\left(\sum\limits_i \left(-\lambda^i_{L}\bar{\psi}^i_{L} \gamma_a  \psi^i_L + \lambda^i_{R}\bar{\psi}^i_{R} \gamma_a \psi^i_{R}\right)\right)^2\,,
\end{equation}
where the sum is over all species of fermions, { and we have also redefined the $\lambda$ by absorbing $\sqrt{\frac{3\kappa}{8}}$\,}. We identify this term as  the \textbf{torsional interaction} term which is usually independent of the background metric, but can modify it through Einstein equations. This results a curvature which is generally small enough so that we can take it as a flat spacetime and do normal QFT calculations. We emphasize that we do not get this term by extending or modifying GR, but this is how ordinary fermions behave in a spacetime which is not flat. Along with this interaction, Standard Model interactions are always present -- both will contribute to the calculations of neutrino mixing.

In this paper, we mainly concern ourselves about how neutrino oscillations will be affected in presence of torsional four-fermion interaction. We generally believe that matter effects suppress the effects of curvature on neutrino oscillations, even in regions of strong gravity such as supernovae~\cite{Cardall:1996cd}. Our approach here is completely different from ordinary gravitational effect because of the dimensionful coupling constants $\lambda^i_{L, R}$ which are not universal but { can be fixed only from experimental observations.}
\section{Two flavors of neutrino}
\label{sec:2flavor}
Let us first begin with two species of neutrinos passing through normal matter of uniform density. It is the field in the mass basis which couple to torsion, since torsion appears with the geometric connection. Interaction of neutrinos with background is
\begin{small}
\begin{equation}\label{L.1}
		-\left(\sum_{i=1,2}\left(-\lambda_{i}^{L}\bar{\nu}_i \gamma_a L\nu_i + \lambda_{i}^R \bar{\nu}_i \gamma_a R \nu_i \right)\right) \times
			\left(\sum_{f=e, p, n}\left(\lambda_{f}^{V}\bar{f} \gamma_a f + \lambda_{f}^A \bar{f} \gamma_a\gamma^5 f \right)\right) 
		\,.
\end{equation}
\end{small}
The sum includes only $e,p,n$ because other fermions have negligible presence  in the atmosphere or the Sun. In most situations the density of matter is not sufficient to cause high curvature, so we can neglect the $\omega_\mu{}^{ab}$ term and do the calculations as in case of a flat background. Like weak interactions, the background factor can be replaced by its average value by considering the forward scattering of neutrinos, 
\begin{equation}\label{avg}
  \sum_{f=e, p, n}\left\langle\lambda_{f}^{V}\bar{f} \gamma_a f + \lambda_{f}^A \bar{f} \gamma_a\gamma^5 f \right\rangle\,
\end{equation}
If the background consists of non-relativistic fermions, the average becomes the number density. Thus the interaction term is 
\begin{equation}\label{L.eff}
 -\left( \sum_{i=1,2}\left(-\lambda_{i}^{L}\bar{\nu_i} \gamma_0 L\nu_i + \lambda_{i}^R \bar{\nu_i} \gamma_0 R \nu_i\right) \right) \tilde{n},
  \end{equation}
Where $ \tilde{n} $ is the weighted number density of the background matter, $\tilde{n} = \sum\lambda^V_{f} n_f.$ We also consider maximal chirality violation, so $ \lambda^R_i=0 $ for neutrinos. Then the contribution to the effective Hamiltonian is
\begin{equation}\label{H.eff}
 \sum_{i=1,2}\left(\lambda_{i}{\nu}_i^\dagger L \nu_i \right)\,\tilde{n}\,.
\end{equation}
The flavor eigenstates $\ket{\nu_\alpha}$ can be written in terms of the mass eigenstates $\ket{\nu_i}$ as
\begin{equation}
 \ket{\nu_{\alpha}}=\sum_{i}U^{*}_{\alpha i}\ket{\nu_{i}}\,,
\end{equation}
where the mixing matrix $U=\begin{pmatrix}\cos\theta & \sin\theta \\ -\sin\theta\, & \cos\theta\end{pmatrix}$\,. Following~\cite{Wolfenstein:1977ue}, we can now write the Schr\"{o}dinger equation for the neutrinos,
%
\begin{align}\label{SE.2nu-mass}
i\frac{d}{dx}\begin{pmatrix}\nu_1\\ \nu_2\\ \end{pmatrix} =&\left[E{\mathbb I} +\frac{1}{2E}\begin{pmatrix} m_1^2 & 0 \\ 0 & m_2^2\\ \end{pmatrix}+\begin{pmatrix}\lambda_1 & 0 \\ 0 & \lambda_2\\ \end{pmatrix} \tilde{n}-\frac{G_F}{\sqrt{2}}(n_n-n_e) \right. \notag\\ 
&\qquad\qquad\left. + \frac{G_F}{\sqrt{2}}U^{T}\begin{pmatrix}n_e & 0 \\ 0 & -n_e\\ \end{pmatrix}U^{*}\right]\begin{pmatrix}\nu_1\\ \nu_2\\ \end{pmatrix},
\end{align}
%
where the effect of weak interaction has also been included.
Let us define a torsionally modified mass-squared difference $ \Delta m_s^2 $ as
\begin{equation}\label{ms-sq}
  \Delta m_s^2 =\Delta m^2+2 \tilde{n}E \Delta \lambda\,,
\end{equation}
where $ \Delta m^2 =m_2^2-m_1^2 $ and $ \Delta\lambda = \lambda_2 - \lambda_1\,. $
It can be shown easily that the mixing angle in matter, modified by the torsional four-fermion interaction, is given by~\cite{Ghose:2023ttq}
\begin{equation}\label{mixing-angle}
 \tan 2\theta_M=\frac{\tan 2\theta}{1-\frac{D}{\Delta m_s^2 \cos 2\theta}}\,,
\end{equation}
where $D = 2\sqrt{2}G_F n_e E\,.$
By diagonalizing Eq.(\ref{SE.2nu-mass}), we can find the $ \nu_e \to \nu_\mu $ conversion probability 
\begin{equation}\label{conversion-2nu}
 P_{\nu_e \to \nu_{\mu}}=\sin ^2(2\theta_M)\sin^2\left(\frac{\Delta m_M^2}{4E}L\right)\,,
\end{equation}
and the $ \nu_e $ survival probability
\begin{equation}\label{survival-2nu}
 P_{\nu_e \to \nu_e}=1-\sin ^2(2\theta_M)\sin^2\left(\frac{\Delta m_M^2}{4E}L\right)\,,
\end{equation}
where for convenience we have written
\begin{equation}\label{delta-mM}
	\Delta m_M^2=\sqrt{(\Delta m_s^2\cos 2\theta-D)^2+(\Delta m_s^2 \sin 2\theta)^2}\,.
\end{equation}
Therefore we see that spacetime geometry modifies the mass squared differences via Eq.~(\ref{ms-sq}) and mixing angle via Eq.~(\ref{mixing-angle})and thus modifies the oscillation probabilities through Eq.~(\ref{conversion-2nu}) and Eq.~(\ref{survival-2nu}).
\section{Three flavors of neutrino}
\label{sec:3flavor}
In nature we have three species of neutrinos. Let us now consider the effect of spacetime geometry, and thus torsional four-fermion interaction, on mixing between three flavors of neutrinos. It is known that if more than two families of neutrino exist, CP and T can be broken via complex elements of mixing matrix~\cite{Kobayashi:1973fv,Cabibbo:1977nk}. We follow the conventions of Particle Data Group (PDG) and write the mixing matrix as~\cite{ParticleDataGroup:2020ssz}
\begin{equation}\label{3nu.mixing}
U=\begin{pmatrix}c_{12}c_{13} & s_{12}c_{13} & s_{13}e^{-i\delta}\\-s_{12}c_{23}-c_{12}s_{23}s_{13}e^{i\delta} & c_{12}c_{23}-s_{12}s_{23}s_{13}e^{i\delta} & s_{23}c_{13}\\ s_{12}s_{23}-c_{12}c_{23}s_{13}e^{i\delta} & -c_{12}s_{23}-s_{12}c_{23}s_{13}e^{i\delta} & c_{23}c_{13}\end{pmatrix}
\end{equation}
where $c_{ij}=\cos\theta_{ij}$ and $s_{ij}=\sin\theta_{ij}\,,$ where $\theta_{ij}$ are the mixing angles, $\delta$ is the CP-violating phase, and we have ignored Majorana phases as we are dealing with Dirac neutrinos. The angles $\theta_{ij}$ are in the first quadrant and the CP-violation phase $\delta$ is taken to be between $0$ and $2\pi$\,. Then $ U $ is conveniently expressed as a product of rotation matrices ${\mathcal O}_{ij}$ for rotation in the $ij$-plane as~\cite{Akhmedov:2004ny}
\begin{equation}\label{U-pdt form}
	U=\mathcal{O}_{23}\mathcal{U}_{\delta}\mathcal{O}_{13}\mathcal{U}_{\delta}^{\dagger}\mathcal{O}_{12}\,, 
\end{equation}
where $U_\delta = \mathrm{diag}(1,\, 1,\, e^{i\delta})\,.$ The 
Schr\"odinger equation is written in the mass basis, similarly to Eq.~\ref{SE.2nu-mass},
%
\begin{align}
i\frac{d}{dx}\begin{pmatrix}{\nu_1}\\ {\nu_2} \\ \nu_3\end{pmatrix}=& \left[E+\frac{1}{2E}\begin{pmatrix}m_1^2 & 0 & 0 \\ 0 & m_2^2 & 0 \\ 0 & 0 & m_3^2\end{pmatrix}+\begin{pmatrix}\lambda_1 & 0 & 0 \\ 0 & \lambda_2 & 0 \\ 0 & 0 & \lambda_3\end{pmatrix}\tilde{n} \right. \notag \\ 
&\qquad \qquad\left. -\frac{G_F}{\sqrt{2}}n_n+U^{T}\begin{pmatrix}A & 0 & 0 \\ 0 & 0 & 0 \\ 0 & 0 & 0\end{pmatrix}U^{*}\right]\begin{pmatrix}\nu_1 \\ \nu_2 \\ \nu_3 \end{pmatrix} \label{SE.3nu-mass}
\end{align}
%
where $A = {\sqrt{2}}{G_F} n_e\,.$
For uniform matter density or slowly varying matter we can write in the flavor basis
\begin{small}
\begin{align}
i\begin{pmatrix}{\dot\nu_e}\\ {\dot\nu_\mu} \\ \dot\nu_\tau \end{pmatrix} &=\left[E'_0\mathbb{I}+\frac{1}{2E}U^*\begin{pmatrix}0 ~& 0 & 0 \\ 0 ~& \Delta \tilde{m}_{21}^2 & 0 \\ 0 ~& 0 & \Delta \tilde{m}_{31}^2\end{pmatrix}U^T +\begin{pmatrix}A & 0 & 0 \\ 0 & 0 & 0 \\ 0 & 0 & 0\end{pmatrix}\right] \begin{pmatrix}\nu_e \\ \nu_\mu \\ \nu_\tau \end{pmatrix} \,.\label{SE.3nu.flavor}
\end{align}
\end{small}
We have used the definition
\begin{align}
E'_0=E+\frac{m_1^2+2\lambda_1\tilde{n}E}{2E}-\frac{G_F}{\sqrt{2}}n_n\,. \nonumber
\end{align}
In order to find the eigenvalues of the resulting Hamiltonian, we take the help of perturbation theory using a small parameter. For this, we first define
\begin{equation}\label{tm-sqd}
	\Delta \tilde{m}^2_{ij} := \Delta m^2_{ij} +2 \tilde{n} E \Delta\lambda_{ij}\,,
\end{equation}
where $\Delta m_{ij}^2=m_i^2-m_j^2$ and $\Delta \lambda_{ij}=\lambda_{i}-\lambda_{j}\,.$ Then Eq.~(\ref{SE.3nu.flavor}) takes the form 
\begin{align}
	i \frac{d}{dt}\begin{pmatrix}{\nu_e}\\ {\nu_\mu} \\ \nu_\tau\end{pmatrix}
	&= \frac{\Delta \tilde{m}_{31}^2}{2E}{\cal O}_{23}{\cal U}_\delta^{*} M {\cal U}_\delta^{T} {\cal O}_{23}^T\,\begin{pmatrix}{\nu_e}\\ {\nu_\mu} \\ \nu_\tau\end{pmatrix} .\label{SE.3nu_short} 
\end{align}
Here the matrices proportional to the identity matrix have been suppressed as they will contribute to a common phase for all neutrinos and thus has no effect on oscillation probabilities.
We have also written 
\begin{equation}\label{3M}
	M={\cal O}_{13}{\cal O}_{12}\begin{pmatrix}0 & 0 & 0\\0 & \alpha & 0\\0 & 0 & 1\end{pmatrix}
	{\cal O}_{12}^T {\cal O}_{13}^T+\begin{pmatrix} \hat{A} & 0 & 0\\0 & 0 & 0\\0 & 0 & 0\end{pmatrix} \,.
\end{equation}
In these expressions, $\hat{A} =2AE/\Delta \tilde{m}_{31}^2$ and $\alpha= \Delta \tilde{m}_{21}^2/\Delta \tilde{m}_{31}^2$ are dimensionless quantities.
Now our main focus is to find out  the eigenvalues and eigenvectors of the resulting Hamiltonian. Exactly diagonalizing a $3\times 3$ matrix is quite difficult and thus { we need to use some approximations at this point~\cite{Akhmedov:2004ny, Nunokawa:2005nx, Minakata:2006gq}. We will assume that $s_{13}$ and $\alpha$\, are small parameters and then find the transition probabilities at second order of these parameters.}

In order to find the mixing matrix, we will diagonalize $M$ by finding the eigenvalues and eigenvectors using perturbation theory.
Let us first decompose $M$ into three parts containing different powers of the small parameters $\alpha$\, and $s_{13}$\,,
\begin{align}
	M^{(0)} &=\begin{pmatrix}\hat{A} & 0 & 0\\ 0 & 0 & 0 \\ 0 & 0 &1\end{pmatrix} \nonumber \\
	M^{(1)} &=\begin{pmatrix}\alpha s_{12}^2 & \alpha s_{12} c_{12} & s_{13}\\ \alpha s_{12}c_{12} & \alpha c_{12}^2 & 0 \\ s_{13} & 0 & 0\end{pmatrix}\nonumber \\
	M^{(2)} &=\begin{pmatrix}s_{13}^2 & 0 & -\alpha s_{13}s_{12}^2\\ 0 & 0 & -\alpha s_{13}s_{12}c_{12}\\ -\alpha s_{13}s_{12}^2 & -\alpha s_{13}s_{12}c_{12} & -s_{13}^2\end{pmatrix}\,. \nonumber
\end{align}
Using perturbation theory, it is straightforward to calculate the eigenvalues $\mu$ and the eigenvectors $v$\,. Keeping terms up to the second order in the small parameters $\alpha$ and $s_{13}$\,,
we find the eigenvalues
\begin{align}
	\mu_1&=\hat{A}+\alpha s_{12}^2+s_{13}^2\frac{\hat{A}}{\hat{A}-1}+\frac{\alpha^2 \sin^2(2\theta_{12})}{4\hat{A}} \label{1st_eigenvalue}\\
	\mu_2&=\alpha c_{12}^2-\frac{\alpha^2 \sin^2(2\theta_{12})}{4\hat{A}} \label{2nd_eigenvalue}\\
	\mu_3&=1-s_{13}^2\frac{\hat{A}}{\hat{A}-1}\,. \label{3rd_eigenvalue}
\end{align}
The Hamiltonian in Eq.~(\ref{SE.3nu.flavor}) is related to $M$ by a unitary transformation as seen in Eq.~(\ref{SE.3nu_short}). Hence, the energy eigenvalues of the Hamiltonian are 
\begin{equation}\label{energy-ev}
	E_i = \frac{\Delta \tilde{m}_{31}^2}{2E} \mu_i\,.
\end{equation}
The zeroth order eigenvectors are the basis vectors $\hat{e}_i = \begin{pmatrix}1\\0\\0\end{pmatrix}\,, \begin{pmatrix}0\\1\\0\end{pmatrix}\,, \begin{pmatrix}0\\0\\1\end{pmatrix}\,.$ 
By using perturbation theory we can calculate the higher order corrections to the eigenvectors as
\begin{align}
	v_i^{(1)}&=\sum_{i \ne j} \frac{M_{ij}^{(1)}}{\mu_i^{(0)}-\mu_{j}^{(0)}}\hat{e}_j \nonumber\\
	v_i^{(2)}&=\sum_{j \ne i}\frac{1}{\mu_{i}^{(0)}-\mu_j^{(0)}}(M_{ij}^{(2)}+(M^{(1)}v_i^{(1)})_j-\mu_i^{(1)}(v_i^{(1)})_j)\hat{e}_j
	\,. \nonumber
\end{align}
We will rewrite Eq.~(\ref{SE.3nu_short}) in terms of the diagonal matrix $\hat{M}$ and the diagonalizing matrix $W$, which diagonalizes $M$ into $\hat{M}$.
\begin{align}
	i\frac{d}{dx} \begin{pmatrix}{\nu_e}\\ {\nu_\mu} \\ \nu_\tau\end{pmatrix} &=\frac{\Delta \tilde{m}_{31}^2}{2E}{\cal O}_{23}{\cal U}_\delta^{*}W\hat{M}W^T{\cal U}_\delta^T{\cal O}_{23}^T \begin{pmatrix}{\nu_e}\\ {\nu_\mu} \\ \nu_\tau\end{pmatrix} \nonumber \\
	&=\frac{\Delta \tilde{m}_{31}^2}{2E}U'^{*}\hat{M}U'^{T} \begin{pmatrix}{\nu_e}\\ {\nu_\mu} \\ \nu_\tau\end{pmatrix} = \frac{\Delta \tilde{m}_{31}^2}{2E}U'^{*}HU'^{T} \begin{pmatrix}{\nu_e}\\ {\nu_\mu} \\ \nu_\tau\end{pmatrix} \,. \label{SE.3nu.2}
\end{align}
Here $U'$, the new mixing matrix, is defined as $U'={\cal O}_{23}{\cal U}_\delta W$, which is calculated to be
\begin{tiny}
\begin{align}
	U'=\begin{pmatrix}1-\frac{1}{2}\frac{\alpha^2}{\hat{A}^2}c_{12}^2s_{12}^2 & -\frac{\alpha}{\hat{A}}s_{12}c_{12}(1+\frac{\alpha}{\hat{A}}\cos(2\theta_{12})) & -\frac{s_{13}}{\hat{A}-1}(1-\frac{\alpha \hat{A}}{\hat{A}-1}s_{12}^2)\\-\frac{1}{2}\frac{s_{13}^2}{(\hat{A}-1)^2} & & \\ \\ \frac{\alpha}{\hat{A}}c_{12}s_{12}(1+\frac{\alpha}{\hat{A}}\cos(2\theta_{12}))c_{23} & (1-\frac{1}{2}\frac{\alpha^2}{\hat{A}^2}s_{12}^2c_{12}^2)c_{23} & -\frac{\alpha s_{13}}{\hat{A}-1}\hat{A}s_{12}c_{12}c_{23}\\+\frac{s_{13}}{\hat{A}-1}(1-\frac{\alpha \hat{A}}{\hat{A}-1}s_{12}^2)e^{i\delta}s_{23} & +\alpha s_{13}s_{12}c_{12}(1+\frac{1}{\hat{A}})e^{i\delta}s_{23} & +(1-\frac{1}{2}\frac{s_{13}^2}{(\hat{A}-1)^2})e^{i\delta}s_{23}\\ \\-\frac{\alpha}{\hat{A}}c_{12}s_{12}(1+\frac{\alpha}{\hat{A}}\cos(2\theta_{12}))s_{23} & -(1-\frac{1}{2}\frac{\alpha^2}{\hat{A}^2}s_{12}^2c_{12}^2)s_{23} & +\frac{\alpha s_{13}}{\hat{A}-1}\hat{A}s_{12}c_{12}s_{23}\\ +\frac{s_{13}}{\hat{A}-1}(1-\frac{\alpha \hat{A}}{\hat{A}-1}s_{12}^2)e^{i\delta}c_{23} & +\alpha s_{13}s_{12}c_{12}(1+\frac{1}{\hat{A}})e^{i\delta}c_{23} & +(1-\frac{1}{2}\frac{s_{13}^2}{(\hat{A}-1)^2})e^{i\delta}c_{23}\end{pmatrix}\,.
\end{align}
\end{tiny}
Using this mixing matrix various transition probabilities can be calculated quite easily. For example, the expression for the amplitude of conversion  of $\nu_e$ to $\nu_{\mu}$ is
%
\begin{align}
	A_{\nu_e \to \nu_{\mu}} &= \bra{\nu_{\mu}}e^{-iHt}\ket{\nu_e} \nonumber \\
	&=\left(\frac{\alpha}{\hat{A}}{s}_{12}{c}_{12}(1+\frac{\alpha}{\hat{A}}\cos(2\theta_{12})){c}_{23} \right. \notag \\
	& \left. +\frac{{s}_{13}}{\hat{A}-1}(1-\frac{\alpha \hat{A}}{\hat{A}-1}{s}_{12}^2)e^{i\delta}{s}_{23}\right)e^{-iE_1t} \nonumber \\
	&+\left(-\frac{\alpha}{\hat{A}}{s}_{12}{c}_{12}(1+\frac{\alpha}{\hat{A}}\cos(2\theta_{12})){c}_{23}\right)e^{-iE_2t} \notag \\ 
	&+\left(-\frac{{s}_{13}}{\hat{A}-1}(1-\frac{\alpha \hat{A}}{\hat{A}-1}s_{12}^2)e^{i\delta}{s}_{23}\right)e^{-iE_3t}\,.
\end{align}
%
The conversion probability is given by $P_{\nu_e \to \nu_{\mu}}=|A_{\nu_e \to \nu_{\mu}}|^2$. 
Other conversion amplitudes (e.g. $\nu_e \to \nu_\tau$) can be similarly calculated, leading to the corresponding conversion probabilities as well as  survival probability ($P_{\nu_e \to \nu_{e}}$). Our result exactly matches with those found in~\cite{Barger:1980tf,Zaglauer:1988gz} if the torsional interaction is set to zero. {The amplitudes and probabilities for $\nu_\mu\to \nu_e\,, \nu_\mu\to\nu_\tau\,,$ and $\nu_\mu\to \nu_\mu$ were shown elsewhere recently~\cite{Barick:2023wxx}, along with the difference from $\lambda = 0$ results.
Fitting these results to neutrino data should produce an estimate of the coupling parameters $\lambda$.

It is important to recognize that it is not possible to obtain bounds on $\lambda$ from purely theoretical considerations.  Since the four-fermion interaction appears from  the spin connection, it is enticing to think of it as fundamentally gravitational and thus expect the couplings to be of size $\sim\frac{1}{M_p}\,.$ This is a red herring. Four-fermion interactions arise in effective gauge theories when the boson propagator $\sim\frac{1}{q^2 - M^2}$ in an exchange process is replaced by $-\frac{1}{M^2}$ in the low energy limit $|q^2| \ll M^2$\,. This is what happens in Fermi's theory of weak interactions, for example. Here, the geometrical interactions do not appear in the low energy limit of any theory -- the contorsion field $\Lambda_\mu^{\,ab}\,,$ which was eliminated to produce these interactions, is not dynamical and does not have a propagator. We have not quantized gravity or spacetime, so the scale of quantum gravity is not relevant here. The couplings should not change even at very high energy scales. When the couplings $\lambda$ were redefined to absorb a factor of $\kappa$\,, it was only for bookkeeping purposes, there was no reason to take them to be small dimensionless numbers before that.
} This torsional interaction is one kind of Non-Standard Interaction (NSI), so known bounds on various NSI coupling parameters should be compared with the parameters which appear here. 

\medskip

* The authors declare that there is no conflict of interest regarding the work reported in this paper.


\end{document}